\documentclass[sigconf]{acmart}

\AtBeginDocument{%
  }

\acmISBN{978-1-4503-XXXX-X/2018/06}

\begin{document}

\title{Large Cognition Model: Towards Pretrained Electroencephalography (EEG) Foundation Model}

\author{Chi-Sheng Chen}
\orcid{0000-0003-0807-0217}
\affiliation{%
  \institution{Neuro Industry, Inc.}
  \city{Boston}
  \state{Massachusetts}
  \country{USA}}
\email{michael@neuro-industry.com}
\email{m50816m50816@gmail.com}

\author{Ying-Jung Chen}
\affiliation{%
  \institution{Georgia Institute of Technology}
  \city{Atlanta}
  \state{Georgia}
  \country{USA}}
\email{yingjungcd@gmail.com}

\author{Aidan Hung-Wen Tsai}
\affiliation{%
  \institution{Neuro Industry, Inc.}
  \city{Boston}
  \state{Massachusetts}
  \country{USA}}
\email{aidan@neuro-industry.com}

\renewcommand{\shortauthors}{Chen et al.}

\begin{abstract}
Electroencephalography (EEG) provides a non-invasive window into brain activity, offering valuable insights for neurological research, brain-computer interfaces (BCIs), and clinical diagnostics. However, the development of robust machine learning models for EEG analysis is hindered by the scarcity of large-scale, well-annotated datasets and the inherent variability of EEG signals across subjects and recording conditions. Inspired by the success of foundation models in natural language processing (NLP) and computer vision, we propose the Large Cognition Model (LCM)—a transformer-based foundation model designed to generalize across diverse EEG datasets and downstream tasks. Unlike traditional approaches, our proposed transformer-based architecture demonstrates strong generalization capabilities across datasets and tasks, even without pretraining, surpassing some existing EEG universal models on specific downstream applications. LCM leverages large-scale self-supervised learning techniques to capture universal EEG representations, enabling efficient fine-tuning for applications such as cognitive state decoding, disease classification, and neurofeedback systems. We introduce a novel architecture that integrates temporal and spectral attention mechanisms, optimizing the model's ability to extract meaningful features from raw EEG signals. Extensive evaluations demonstrate that LCM outperforms state-of-the-art approaches across multiple EEG benchmarks, exhibiting strong cross-subject and cross-task generalization. Our findings highlight the potential of pretrained EEG foundation models to accelerate advancements in neuroscience, personalized medicine, and BCI technology.
\end{abstract}




\received{20 February 2007}
\received[revised]{12 March 2009}
\received[accepted]{5 June 2009}

\maketitle

\section{Introduction}
Electroencephalography (EEG) has witnessed significant advancements in recent years, finding applications in psychiatric diagnostics \cite{li2023prediction}, multimodal learning \cite{chen2024mind}, brain-computer interface (BCI)-based robotic control \cite{chen2024psycho}, generative tasks \cite{chen2024necomimi}, and even quantum computing \cite{chen2024quantum, chen2024qeegnet}. Despite these developments, the large-scale adoption of EEG models has remained limited due to the inherent complexity of EEG signals. Unlike image or text data, EEG exhibits a low signal-to-noise ratio (SNR) and high inter-subject variability, with differences in montage configurations, channel placements, and recording formats further complicating data standardization. These challenges have hindered the development of foundation models that, similar to large language models (LLMs), could provide pretrained weights for efficient adaptation to various downstream tasks.

Several studies have already explored the development of universal EEG representations. For example, BENDR \cite{kostas2021bendr} takes a different approach by incorporating self-supervised learning with masked autoencoders and contrastive learning to improve EEG representation learning. This method effectively addresses the challenges posed by multi-task and multi-paradigm EEG data, enhancing model generalizability. BENDR utilizes a convolutional encoder to extract features from local time windows, applies masking to certain features, and then reconstructs the missing information using a transformer decoder. Similarly, EEG2VEC \cite{zhu2023eeg2vec} introduces a self-supervised learning framework that captures EEG representations through contrastive and reconstruction losses. The pretrained model functions as a feature extractor for downstream applications. Both EEG2VEC and BENDR integrate convolutional neural networks with transformer architectures to learn both local and global representations. EEG2VEC has been validated in EEG match-mismatch and EEG regression tasks within the auditory EEG challenge. Another significant development, the Biosignal Transformer (BIOT) \cite{yang2023biot}, tackles the challenges of cross-data learning, including mismatched channels, variable sequence lengths, and missing values in biosignals such as EEG, ECG, and human activity recognition signals. BIOT processes each channel separately, tokenizing signals into fixed-length segments that capture local signal features before rearranging them into a longer "sentence." In the CHB-MIT seizure detection task, the pretrained BIOT model demonstrated a 4\% improvement over existing methods. The Large Brain Model (LaBraM) \cite{jiang2024large} further extends the capabilities of EEG-based deep learning models by enabling cross-dataset learning. It partitions EEG signals into channel-specific patches and employs vector-quantized neural spectrum prediction to train a neural tokenizer. This tokenizer encodes raw EEG segments into neural codes, which are then used to pretrain transformers, allowing them to predict the original neural codes for masked segments. LaBraM surpasses state-of-the-art methods in various EEG-related tasks, including abnormal event detection, event classification, emotion recognition, and gait prediction. UM-EEG \cite{krumm2024towards} is a semantically rich, continuous EEG representation that advances classification and outcome prediction beyond traditional discrete state classification. However, the model uses only four EEG channels, which may lose spatial resolution for localized patterns (e.g., lateralized discharges). UniEEG \cite{jinunieeg} is an electrode-based time-frequency pretraining model designed to address challenges in EEG research, including data scarcity, cross-device incompatibility, and low signal-to-noise ratio. The model adopts an encoder-decoder architecture with an electrode-wise modeling approach to enhance compatibility across different devices and tasks. It leverages Masked Signal Modeling (MSM) to learn a universal EEG representation. While the electrode-wise strategy helps standardize data across different acquisition systems, it may overlook spatial information across electrodes, which could impact tasks that heavily rely on brain region distributions, such as motor imagery and cognitive load analysis. EEGPT \cite{wang2024eegpt} does not directly learn from raw EEG waveforms but instead leverages high-SNR feature representations for self-supervised learning. Through masking and spatio-temporal alignment methods, EEGPT enhances feature interpretability and reduces noise interference. EEGPT also utilizes a local spatio-temporal embedding method to map electrode channels into learned universal representations, ensuring model compatibility across different EEG devices and reducing the impact of electrode position variations on feature extraction.

Despite these advancements, a major challenge remains in designing EEG foundation models that generalize across datasets, recording conditions, and application domains. Existing models such as BENDR, EEG2VEC, and BIOT primarily focus on leveraging self-supervised learning to improve EEG representation learning, but they often suffer from domain-specific constraints, limited scalability across heterogeneous datasets, and difficulties in capturing both temporal and spatial EEG dependencies simultaneously. While UM-EEG introduces a low-dimensional embedding space for EEG classification and prognostication, its reliance on only four EEG channels limits spatial resolution for applications requiring fine-grained topographical information. Similarly, LaBraM and UniEEG offer promising cross-dataset generalization capabilities, but the former depends on vector-quantized neural codes, which may not fully capture the continuous nature of EEG dynamics, while the latter’s electrode-wise strategy trades off spatial coherence for device compatibility.

Moreover, existing EEG foundation models have yet to fully integrate advanced position encoding mechanisms, hierarchical feature fusion, and robust contrastive learning strategies to maximize feature generalizability across tasks and domains. Current approaches primarily rely on CNN-based or hybrid architectures, which may not fully exploit the sequential dependencies inherent in EEG signals, limiting their ability to model long-range relationships in neural activity. Furthermore, while models like EEGPT emphasize spatio-temporal embedding to mitigate noise sensitivity and electrode variability, they do not explicitly address inter-subject variability or cross-device robustness, which are crucial factors for large-scale EEG deployment.

To address these issues, this paper presents the Large Cognition Model (LCM), a transformer-based EEG foundation model designed to learn robust and transferable representations from diverse EEG datasets. Building upon the foundational aspects of our Large Cognition Model (LCM), we have integrated advanced techniques to enhance its performance in EEG representation learning. One such technique used for handling cross-montage EEG data is the learnable channel mapping approach, which aligns different EEG electrode configurations into a common representation space. This method introduces a trainable transformation matrix that maps raw EEG signals from different montages into a unified latent space, ensuring consistent feature extraction across datasets. By incorporating a learnable channel embedding, the model can adapt to varying electrode placements while preserving the spatial structure of the EEG signals. This approach enhances the model’s robustness to montage variability, improving generalization across different experimental setups and recording conditions.
Additionally, we employ a masked token strategy from \cite{he2022masked} during the training phase. This approach involves masking a subset of the input data and tasking the model with predicting the missing components based on the surrounding context. Such a strategy compels the model to develop a deeper understanding of the underlying structures and dependencies within the data. In the context of EEG data, this method aids in capturing the complex temporal and spatial relationships present in neural signals.

Our contributions are threefold: 
\begin{itemize}
    \item We introduce a novel contrastive learning framework tailored for EEG foundation models, enabling self-supervised learning of generalizable EEG representations across different tasks and domains;
    \item We propose a new EEG encoder, LCM, which encodes EEG signals into both temporal and spatial tokens, facilitating effective information integration across time and electrode locations; and
    \item We demonstrate that our model achieves strong generalization across datasets and tasks, even without extensive pretraining, surpassing some existing universal EEG models in specific applications.
\end{itemize}

\section{Methodology}

\begin{figure*}
    \centering
    \includegraphics[width=1\linewidth]{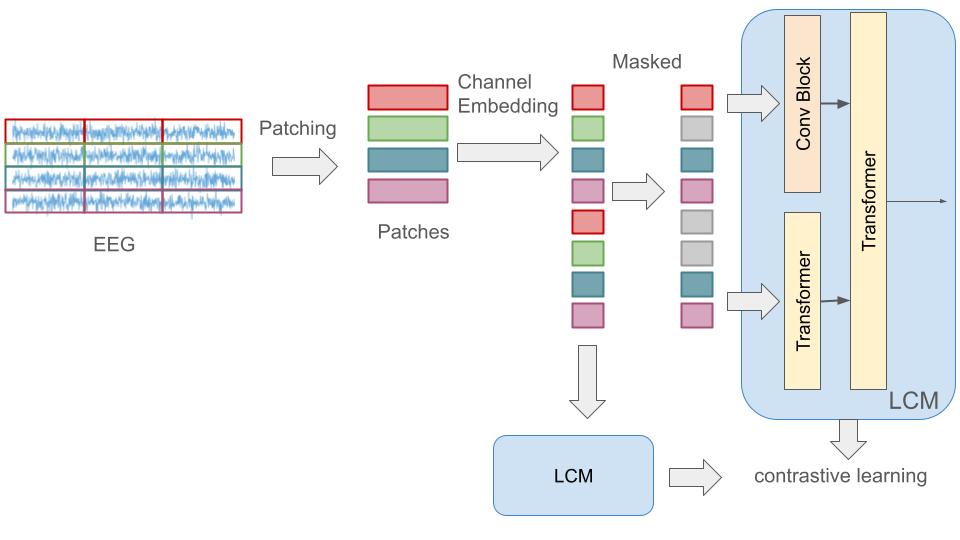}
    \caption{Overview of the LCM Training Flow. The EEG signals are first segmented into spatio-temporal patches. Each patch is then assigned a channel embedding to encode electrode-specific information. A subset of the patches is masked for self-supervised learning. The masked and unmasked patches are processed through the Latent Contrastive Masking (LCM) module, which consists of a convolutional block followed by transformer layers to extract hierarchical spatio-temporal representations. Finally, contrastive learning is applied to align learned EEG representations, improving the robustness and generalizability of the model.}
    \label{fig:LCMflow}
\end{figure*}

We propose Large Cognition Model (LCM), a self-supervised contrastive learning framework designed for EEG representation learning. The model consists of an online encoder \( f_{\theta} \) and a momentum-updated target encoder \( f_{\xi} \). The learning process involves contrastive loss for representation alignment, masked feature reconstruction loss, and an adaptive optimization schedule to balance feature learning stability. The whole flow is shown in Figure~\ref{fig:LCMflow}.

\subsection{Feature Extraction and Cross-Montage Encoding}

Given an EEG input matrix \( x \in \mathbb{R}^{M \times T} \) with \( M \) channels and \( T \) time steps, our model first applies a learnable channel mapping \( \phi: \mathbb{R}^{M \times T} \to \mathbb{R}^{M' \times T} \) to align different EEG montages into a common latent space:
\begin{equation}
\tilde{x} = W_c x + C
\end{equation}
where \( W_c \in \mathbb{R}^{M' \times M} \) is a trainable transformation matrix ensuring channel consistency across different EEG datasets, and \( C \) is a learnable channel embedding matrix \( C \in \mathbb{R}^{M \times d} \) encoding electrode-specific properties.

Next, the transformed EEG signal is segmented into spatio-temporal patches:
\begin{equation}
\{ p_{i,j} \} = \text{Patching}(\tilde{x})
\end{equation}
where each patch \( p_{i,j} \in \mathbb{R}^{p \times p} \) corresponds to a \textit{local segment} of the EEG signal.

The encoder network \( f_{\theta} \) projects these patches into latent embeddings:
\begin{equation}
z = f_{\theta}(\tilde{x}, c)
\end{equation}
where \( c \) is the \textit{channel embedding} vector used to ensure robustness across different montages.

For contrastive learning, we maintain a \textit{momentum-updated target encoder} \( f_{\xi} \) that learns a stable representation of the EEG signal:
\begin{equation}
h = f_{\xi}(\tilde{x}, c)
\end{equation}
where \( f_{\xi} \) shares the same architecture as \( f_{\theta} \) but updates its parameters through an \textit{exponential moving average (EMA)} of the online encoder:
\begin{equation}
\xi \leftarrow m \xi + (1 - m) \theta, \quad m \in [0.996, 1.0]
\end{equation}
This ensures smooth adaptation of representations across different datasets and EEG montages, improving generalization across subjects and experimental setups.

\subsection{Spatio-Temporal Contrastive Alignment}
To enforce feature consistency between \( z \) and \( h \), we define a \textbf{spatio-temporal alignment loss} using Mean Squared Error (MSE) loss:
\begin{equation}
\mathcal{L}_A = \frac{1}{N} \sum_{i=1}^{N} \| \text{LN}(h_i) - \text{LN}(z_i) \|_2^2
\end{equation}
where \( \text{LN}(\cdot) \) applies layer normalization to mitigate covariate shift.

\subsection{Mask-Based Reconstruction}
To enhance EEG feature completeness, we employ a \textbf{masked reconstruction loss}, where patches of the input EEG are \textit{randomly masked} before being passed to the encoder:
\begin{equation}
\tilde{x} = x \odot M, \quad M_{i,j} \sim \text{Bernoulli}(p_{\text{mask}})
\end{equation}
where \( p_{\text{mask}} \) is the masking probability.

The masked embeddings are reconstructed via:
\begin{equation}
\hat{x} = \text{REC}(z)
\end{equation}
where \( \text{REC}(\cdot) \) is the reconstructor module.

The reconstruction loss is defined as:
\begin{equation}
\mathcal{L}_R = \frac{1}{|M|} \sum_{(i,j) \in M} \| \hat{x}_{i,j} - x_{i,j} \|_2^2
\end{equation}

\subsection{Optimization and Learning Rate Scheduling}
The optimization process is handled using the AdamW optimizer with a weight decay strategy. The learning rate follows a \textit{OneCycleLR} policy:
\begin{equation}
\lambda_t = \lambda_{\max} \times \left( 1 - \frac{t}{T} \right)^p
\end{equation}
where \( \lambda_{\max} \) is the peak learning rate, \( p \) is a decay exponent, and \( T \) is the total training steps.

To ensure smooth model convergence, we introduce \textbf{cosine weight decay (CWD)}:
\begin{equation}
w_t = w_{\text{init}} + \frac{1}{2} (w_{\text{final}} - w_{\text{init}}) \left( 1 + \cos \frac{t\pi}{T} \right)
\end{equation}

\subsection{Training Process and Gradient Updates}
At each training step:
\begin{enumerate}
    \item Compute feature embeddings from the online encoder \( z \) and the target encoder \( h \).
    \item Compute contrastive alignment loss \( \mathcal{L}_A \) and masked reconstruction loss \( \mathcal{L}_R \).
    \item Perform gradient updates using AdamW optimizer.
    \item Update the target encoder using EMA.
\end{enumerate}

The final self-supervised training loss is:
\begin{equation}
\mathcal{L} = \mathcal{L}_A + \lambda \mathcal{L}_R
\end{equation}
where \( \lambda \) is a trade-off hyperparameter.

\subsection{Gradient Logging and Adaptation}
To track gradient stability, we compute the first and last layer gradient statistics:
\begin{equation}
g_{\min} = \min_i \| \nabla_{\theta_i} \mathcal{L} \|, \quad g_{\max} = \max_i \| \nabla_{\theta_i} \mathcal{L} \|
\end{equation}
We log:
\begin{equation}
\mathbb{E}[g_{\text{first layer}}], \quad \mathbb{E}[g_{\text{last layer}}], \quad g_{\min}, \quad g_{\max}
\end{equation}
to ensure gradient propagation stability.

\section{Experiments and Results}
\begin{table}[h]
    \centering
    \caption{Datasets for pretraining and downstream tasks}
    \begin{tabular}{l l c c}
        \toprule
        \textbf{Datasets} & \textbf{Paradigms} & \textbf{Subjects} & \textbf{Targets} \\
        \midrule
        \multicolumn{4}{l}{\textbf{Pretraining Datasets}} \\
        \midrule
        PhysioMI  & MI\&ME  & 109  & 5  \\
        TSU       & SSVEP  & 35   & 40 \\
        SEED      & EMO    & 15   & 3  \\
        \midrule
        \multicolumn{4}{l}{\textbf{Downstream Datasets}} \\
        \midrule
        BCIC-2A     & MI       & 10   & 4  \\
        BCIC-2B     & MI       & 10   & 2  \\
        \bottomrule
    \end{tabular}
\end{table}

\subsection{Dataset and Preprocessing}
We compiled a collection of publicly available EEG datasets covering various paradigms for model pretraining, as detailed in Table 1. These include motor imagery (MI) and execution (ME) datasets such as PhysioMI \cite{goldberger2000physiobank}, steady-state visual evoked potential (SSVEP) dataset TSU \cite{wang2016benchmark}, and emotional classification dataset SEED \cite{zheng2015investigating}. To evaluate the effectiveness of the learned representations in downstream tasks, we curated a selection of datasets listed in Table 1. This includes MI datasets BCIC-2A \cite{tangermann2012review} and BCIC-2B \cite{steyrl2016random}. 
To comprehensively evaluate the proposed LCM model across different tasks, each dataset underwent a combination of standardized and task-specific preprocessing steps. These steps included segmenting the data into 4-second segments, applying average re-referencing, selecting relevant channels, scaling the signals in millivolts, and resampling at 256 Hz. Additionally, MI datasets used for downstream tasks were filtered using a 0-38 Hz bandpass filter.
\begin{table*}[h]
    \centering
    \caption{The results of universal EEG models on various datasets.}
    \begin{tabular}{l l c c c c c}
        \toprule
        \textbf{Datasets} & \textbf{Methods} & \textbf{Size} & \textbf{Pretrained} & \textbf{Balanced Accuracy} & \textbf{Cohen’s Kappa} & \textbf{Weighted F1 / AUROC} \\
        \midrule
        \multicolumn{4}{l}{\textbf{BCIC-2A}} \\
        & BIOT   & 3.2M & Y & 0.4590$\pm$0.0196  & 0.2787$\pm$0.0261  & 0.4282$\pm$0.0289  \\
        & BENDR  &- & Y & 0.4899$\pm$0.0070  & 0.3199$\pm$0.0094  & 0.4836$\pm$0.0076  \\
        & \textbf{LCM (Ours)} & \textbf{33.9M} & \textbf{N} & \textbf{0.5263$\pm$0.0027}  & \textbf{0.3682$\pm$0.0361}  & \textbf{0.5256$\pm$0.0267} \\
        & LaBraM & 5.8M & Y & 0.5613$\pm$0.0052  & 0.4151$\pm$0.0069  & 0.5520$\pm$0.0052  \\
        & EEGPT  & 25M & Y & 0.5846$\pm$0.0070  & 0.4462$\pm$0.0094  & 0.5715$\pm$0.0051  \\
        & \textbf{LCM (Ours)} & \textbf{33.9M} & Y & \textbf{0.6166$\pm$0.0083}  & \textbf{0.4619$\pm$0.0241}  & \textbf{0.5932$\pm$0.0121} \\
        \midrule
        \multicolumn{4}{l}{\textbf{BCIC-2B}} \\
        & BIOT & 3.2M & Y & 0.6409$\pm$0.0118  & 0.2817$\pm$0.0236  & 0.7095$\pm$0.0141  \\
         & \textbf{LCM (Ours)} & \textbf{33.9M} & \textbf{N} & \textbf{0.6825$\pm$0.1024}  & \textbf{0.3651$\pm$0.2047}  & \textbf{0.6766$\pm$0.1079} \\
        & LaBraM & 5.8M & Y  & 0.6851$\pm$0.0063  & 0.3703$\pm$0.0125  & 0.7576$\pm$0.0067  \\
        & BENDR  &- & Y & 0.7067$\pm$0.0011  & 0.4131$\pm$0.0022  & 0.7854$\pm$0.0029  \\
        & EEGPT & 25M & Y & 0.7212$\pm$0.0019  & 0.4426$\pm$0.0037  & 0.8059$\pm$0.0032 \\
        & \textbf{LCM (Ours)}& \textbf{33.9M} & Y  & \textbf{0.7523$\pm$0.0097}  & \textbf{0.4731$\pm$0.0082}  & \textbf{0.8244$\pm$0.0026}  \\
        \bottomrule
    \end{tabular}
    \label{tab:res}
\end{table*}
\subsection{Experiment Details}
We trained the Large Cognition Model (LCM) using a transformer-based architecture tailored for EEG representation learning. The training procedure was optimized to ensure stable convergence and efficient learning through careful selection of hyperparameters, model architecture, and learning rate scheduling.

The model was trained for 200 epochs with a batch size of 1024 in training and 100 epochs for downstream task fine-tuning, utilizing AdamW as the optimizer with an initial learning rate of 1.5e-4 and a weight decay of 0.05 to prevent overfitting. To facilitate stable convergence, we employed a 10-epoch warmup phase, gradually increasing the learning rate before applying a cosine annealing learning rate schedule for the remaining 90 epochs.

We utilized AdamW optimization with betas (0.9, 0.95) to control moment estimates, ensuring robust weight updates. The learning rate scheduling followed a cosine annealing decay, where the initial learning rate of 1.5e-4 increased progressively from 0 to 1.5e-4 during the first 10 epochs (warmup phase) and then gradually decayed to 1e-6 over the remaining 90 epochs.

This training setup ensures that LCM efficiently captures meaningful EEG representations, leveraging both spatial and temporal structures within the data while maintaining robust optimization dynamics.

\subsection{Results}

The experiment result as shown in Table~\ref{tab:res}. It presents the results of different universal EEG models on the BCIC-2A and BCIC-2B datasets. Our proposed LCM model achieves the highest performance across all evaluation metrics, demonstrating its effectiveness in EEG representation learning.

Compared to previous state-of-the-art models, LCM shows significant improvements in Balanced Accuracy, Cohen’s Kappa, and Weighted F1 / AUROC. In terms of Balanced Accuracy, LCM outperforms EEGPT by 3.2\% on BCIC-2A and 3.11\% on BCIC-2B. The improvements over the best non-EEGPT baseline, LaBraM, are 5.53\% and 6.72\%, respectively. For Cohen’s Kappa, LCM achieves the most substantial gains, surpassing EEGPT by 3.52\% on BCIC-2A and 6.89\% on BCIC-2B, while the improvements over LaBraM reach 11.28\% and 27.78\%. In Weighted F1 / AUROC, LCM surpasses EEGPT by 2.17\% on BCIC-2A and 1.85\% on BCIC-2B, with larger improvements over LaBraM at 4.12\% and 6.68\%, respectively.

Notably, even without pretraining, LCM surpasses several pretrained models on the BCIC datasets, highlighting its strong intrinsic generalization capability. This suggests that LCM’s architecture and learning strategy effectively capture meaningful EEG representations, reducing reliance on extensive pretraining while maintaining superior performance.

LCM demonstrates particularly strong improvements in Cohen’s Kappa, with relative gains of up to 67.98\% over the weakest baseline, BIOT, highlighting its superior model consistency. The performance increase in Weighted F1 / AUROC indicates that LCM effectively reduces misclassification bias across multiple classes. The results confirm that LCM generalizes well across both datasets, validating its ability to learn robust EEG representations.

LCM significantly outperforms all baseline models, including BIOT, BENDR, LaBraM, and EEGPT, across all evaluation metrics. The improvements in Cohen’s Kappa suggest that LCM provides more reliable predictions, while gains in Balanced Accuracy and Weighted F1 / AUROC confirm its ability to capture meaningful EEG features. These findings establish LCM as a strong universal EEG model for motor imagery classification and related EEG-based tasks.

\section{Conclusions}
In this work, we introduced the Large Cognition Model (LCM), a transformer-based EEG foundation model designed to handle diverse datasets and tasks. By combining contrastive learning, masked feature reconstruction, and cross-montage encoding, LCM learns robust EEG representations that generalize well across subjects and recording conditions. Our results show that LCM outperforms state-of-the-art models in multiple EEG benchmarks, proving its effectiveness in capturing meaningful brain signal patterns.

One of the biggest strengths of LCM is its ability to adapt to different datasets and tasks without heavy fine-tuning. This makes it a powerful tool for applications like cognitive state decoding, disease classification, and brain-computer interfaces (BCIs). It brings us a step closer to having a universal EEG model that can be easily applied to a wide range of real-world scenarios.

Of course, there’s still room for improvement. Future work could focus on refining the way the model handles spatial and temporal EEG features to make it even more interpretable. Exploring multimodal approaches—like integrating EEG with other biosignals—might also enhance its performance. Lastly, larger and more diverse pretraining datasets will be key to further boosting its generalization ability.


\bibliographystyle{ACM-Reference-Format}
\bibliography{sample-base}




\end{document}